\documentclass[dvips]{article}

\usepackage{graphicx}   % for including figures
\usepackage{amsmath,amssymb}
\usepackage{cite}
\usepackage{xspace}
\usepackage{psfrag}
\usepackage{lineno}
\usepackage{icrc2011}
\newcommand{\semhl}{\ensuremath{S_\mathrm{em,\,halo}}}

\newcommand{\semm}[1]{\ensuremath{S_\mathrm{em}^{\mathrm{\scriptscriptstyle
      #1}}}}
\newcommand{\smum}[1]{\ensuremath{S_\mu^{\mathrm{\scriptscriptstyle
      #1}}}}
\newcommand{\smillam}[1]{\ensuremath{S_{1000}^{\mathrm{\scriptscriptstyle
      #1}}}}

\newcommand{\gsize}{0.42\textwidth}

\newcommand{\gss}{0.7}
\newcommand{\gs}{0.5}
\newcommand{\erange}[2]{\ensuremath{10^{#1}-10^{#2}}~eV}
\newcommand{\arange}[2]{\ensuremath{#1^\circ-#2^\circ}}
\newcommand{\xmax}{\ensuremath{X_\mathrm{max}}}
\newcommand{\xmaxv}{\ensuremath{X_\mathrm{max}^\mathrm{v}}}
\newcommand{\gsm}{g/cm${}^2$}

\newcommand{\sigrat}{\ensuremath{S_\mu/S_\mathrm{em}}}
\newcommand{\sem}{\ensuremath{S_\mathrm{em}}}
\newcommand{\smu}{\ensuremath{S_\mu}}

\newcommand{\logen}{\lg(E/\mathrm{eV})=}

\psfrag{Xmax}[rb][rb][\gss]{\xmaxv, \gsm}
\psfrag{Xmaxv}[rb][rb][\gss]{\xmaxv, \gsm}
\psfrag{SmuSem}[rc][rc][\gss]{$\mathbf{\sigrat}$}
\psfrag{HaloFraction}[rb][rc][\gss]{$\mathbf{\semhl/\sem}$}
\psfrag{SemHaloSmu}[rb][rc][\gss]{$\mathbf{\semhl/\smu}$}
\psfrag{SemHaloFraction}[r][r][\gss]{$\mathbf{\semhl/\sem}$}
\psfrag{Zth*180/TMath::Pi\(\)}[r][r][\gss]{\bf{Zenith angle}}
\psfrag{Zenith angle}[r][r][\gss]{\bf{Zenith angle}}
\psfrag{Smu[3][3]/(Sem[3][3]+Sgm[3][3])}[rc][rc][\gss]{$\mathbf{\sigrat}$}
\psfrag{Smu[3][3]/\(Sem[3][3]+Sgm[3][3]\)/pow\(cos\(Zth\),\2551.2\)}[rb][rb][\gss]{$\sigrat\cos^{1.2}(\theta)$}
\psfrag{Sem33Sgm33Epr1e9143cosZth49cosZthcosZth}[r][r][\gss]{$\sem/(E\cdot f(\theta))$, VEM/EeV}
\psfrag{RSem33Sgm33Epr1e9143cosZth49cosZthcosZth}[r][r][\gss]{$\sem^{\tiny\rm
    EPOS}/\sem^{\tiny\rm QGSJET}$}
\psfrag{Rem33Sgm33Epr1e9143cosZth49cosZthcosZth}[r][r][\gss]{$\sem^{\tiny\rm
    EPOS,\,fit}/\sem^{\tiny\rm QGSJET}$}
\psfrag{126Smu33Epr1e9powcosZth1}[rb][rc][\gss]{$\smu/(E\cdot
  \cos(\theta))$, VEM/EeV}
\psfrag{Energy}[rc][rb][\gss]{$\lg(E/\mathrm{eV})$}
\psfrag{Sem33Sgm33emQGSSem33Sgm33100}[r][r][\gss]{$(\sem^\mathrm{MC}-\sem^\mathrm{fit})/\sem^\mathrm{MC}$, \%}
\psfrag{Smu33muQGSSmu33100}[r][r][\gss]{$(S_\mu^\mathrm{MC}-S_\mu^\mathrm{fit})/S_\mu^\mathrm{MC}$,
  \%}
\psfrag{XmaxS1000_QGS}[b][b][\gs]{$\mathbf{\mathsf{18.9-19.1, 45^\circ-65^\circ}}$}
\title{Applications of $\mathbf\sigrat$ showers universality for mass
  composition and hadronic interactions studies}

\newcommand{\etal}{\MakeLowercase{\textit{et al. }}} % "et al."
\shorttitle{D.~D'Urso \etal Applications of \sigrat\ showers universality}

\authors{D.~D'Urso$^{1}$, M.~Ambrosio$^{1}$, C.~Aramo$^{1}$, M.~Cilmo$^{1,2}$, F.~Guarino$^{1,2}$, L.~Valore$^{1}$, A.~Yushkov$^{3}$}
\afiliations{$^1$INFN Sezione di Napoli, via Cintia, Napoli, Italia 80125\\ 
$^2$Universit\`{a} di Napoli ``Federico~II'', via Cintia, Napoli, Italia 80125\\
$^3$Universidade de Santiago de Compostela, Santiago de Compostela, Espa\~na 15782}
\email{durso@na.infn.it}

\abstract{We present the first results of the application of the
  recently found universality of behavior of muon signal \smu\ to
  electromagnetic (EM) signal \sem\ ratio with respect to the vertical
  depth of showers maximum \xmaxv\ for mass composition and hadronic
  interaction studies. Making use of the fact that for zenith angles
  $>45$ degrees the dependence of \sigrat\ on \xmaxv\ is very similar
  for QGSJET~II and EPOS~1.99 we show that this provides the
  possibility to estimate muon shower content in almost interaction
  model independent way. To evaluate the excess of signal in the data
  in respect to Monte-Carlo predictions we propose to use mass
  independence of the electromagnetic signal. Using the simulations
  with EPOS~1.99 as a fake data we show that one can determine the
  absolute scaling factor between these fake data and the interaction
  model under test (QGSJET~II in our case). Applying this scaling
  factor to the total and muon signals of QGSJET~II one can make
  accurate conclusions on the primary mass of samples prepared with
  EPOS~1.99.}

\begin{document}
\maketitle

\section*{Introduction}
In this paper we apply for mass composition and hadronic interaction
studies the recently found shower universality property, stating that
the ratio of the muon signal to the EM one \sigrat\ is the same for
all hadronic showers having the maximum at the same vertical depth
\xmaxv~\cite{ya_PRD2010,ya_icrc2011}. This property provides a very
simple parametrization for the muon signal~\cite{ya_PRD2010}
\begin{equation}
\label{eq:mufit}
S_\mu^\mathrm{fit}=S_{1000}/(1+1/(\left(\xmaxv/A\right)^{1/b}-a)),
\end{equation}
where $A,\,a$ and $b$ are the fit parameters and $S_{1000}$ is the
total ground plane signal in water Cherenkov detectors similar to the
detectors of the Pierre Auger
Observatory~\cite{PAO_proto_NIMA2004}. We use the large set of CORSIKA
showers for interaction models QGSJET~II/Fluka and EPOS~1.99/Fluka,
described in detail in~\cite{ya_icrc2011}, to demonstrate that for
zenith angles above 45 degrees, where EM halo from muon decays and
interactions composes large part of the EM signal, the discussed
universality allows to find the muon signal in almost interaction
model independent way. This is done using QGSJET~II muon signal
parametrizations on EPOS~1.99 simulations which serve in the given
case as a fake data. Once the muon signal is found the independence of
the \sigrat\ on interaction model properties and independence of the
EM signal on the primary mass are used to determine the scaling factor
between QGSJET~II and EPOS~1.99. It is shown that such scaling
procedure allows to extract with a good accuracy primary mass
composition from EPOS~1.99 samples with the use of the QGSJET~II
model.

\section{$\mathbf{\sigrat}$ universality in inclined
  $\mathbf{\theta>45^\circ}$ showers}

In~\cite{ya_PRD2010} it was demonstrated that the parametrization in
the form~(\ref{eq:mufit}) provides unbiased estimate of the muon
signal with RMS of 8\% and 5\% for protons and iron correspondingly if
the \arange{0}{65} angular range is considered. Here we would like to
study in more detail only inclined $\theta>45^\circ$ showers, since
\sigrat\ here should be very similar for both QGSJET~II and
EPOS~1.99~\cite{ya_icrc2011}.

\begin{figure}[b]
\centering\includegraphics[width=\gsize]{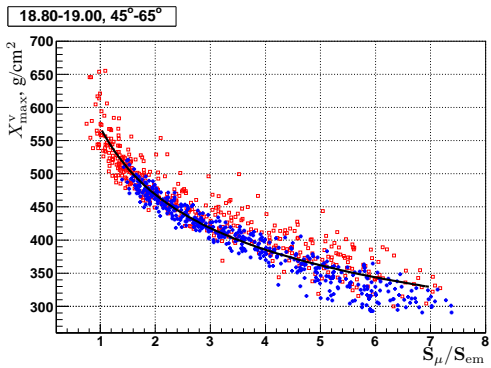}
\caption{\sigrat\ vs \xmaxv\ universality for QGSJET~II. Red
  squares~---~protons, blue crosses~--~iron nuclei, black
  line~---~fit~(\ref{eq:mufit}). $\logen18.8-19.0,\ \theta=45^\circ-65^\circ$.}
\label{fig:univ45}
\end{figure}

From Fig.~\ref{fig:univ45} one can note that \sigrat\ in proton
showers at some fixed \xmaxv\ is slightly larger than in iron ones for
\arange{45}{65} angular range. As discussed in~\cite{ya_PRD2010}
proton showers having the same average depth of maximum \xmaxv\ with
iron showers should be more inclined since they have deeper \xmax. This
in turn means that particles in proton showers should cross larger
slant distance along shower axis from the shower maximum to the
observation level and the fraction of $\pi^0$ EM component in them
should be smaller than in iron showers, bringing to larger \sigrat.
For this reason, as our calculations show, muon signal for $p$ and
Fe showers obtained with the fit~(\ref{eq:mufit}) on average
slightly ($<2\%$) differ from the simulated signals, and for the EM
signals this difference is within 7\% with RMS deviations from the
simulated signal reaching 15\% and 12\% for $p$ and Fe
correspondingly. It is possible to improve the muon and EM signals
recovery. Taking into account that \sigrat\ in proton showers at fixed
\xmaxv\ is larger than in iron showers and that at the same time
proton showers are slightly more inclined than iron ones it is enough
to multiply \sigrat\ by $\cos^\alpha(\theta)$ finding $\alpha$ which
allows to reduce to minimum the difference in
$\sigrat\cos^\alpha(\theta)$ between $p$ and Fe showers. We have found
that this is achieved for $\alpha\approx1-1.5$ and eventually we have
chosen $\alpha=1.2$. In this case the muon signal parametrization
looks like this
\begin{equation}
\label{eq:smucos}
S_\mu^\mathrm{fit}=\frac{S_{1000}}{1+\cos^\alpha(\theta)/\left((\xmaxv/A)^{1/b}-a\right)}.
\end{equation}
we have performed the fits with~(\ref{eq:smucos}) in three energy
ranges $\logen18.5-19.0,\ 19.0-19.5,\ 19.5-20.0$ and the fit
parameters are given in the Table~\ref{tab:fit}. The use of this
approach brings to unbiased estimates of both muon and EM signals with
RMS around 10\% for EM signals and 5\% (3\%) for muon
signal in proton (iron) showers.

\begin{table}[b]
\caption{Fit parameters in~(\ref{eq:smucos}) for QGSJET~II and
  EPOS~1.99 interaction models.}
\label{tab:fit}
\begin{center}
\renewcommand{\tabcolsep}{6pt}
\small
\begin{tabular}{crrrrrr}
\hline
&\multicolumn{3}{c}{QGSJET~II}&\multicolumn{3}{c}{EPOS~1.99}\\
$\lg(E/\mathrm{eV})$ & $A$ & $a$ & $b$ &  $A$ & $a$ & $b$ \\
\hline
$18.5-19.0$ & 2070 &    2.54 &   -1.13  & 10550  &   4.53  &  -1.76\\
$19.0-19.5$ & 1045 &    1.52 &   -0.81  &  2028  &   2.42  &  -1.11\\
$19.5-20.0$ & 742 &     0.94 &   -0.62  &   871  &   1.11  &  -0.70\\
\hline
\end{tabular}
\end{center}
\end{table}

\begin{figure}[thb]
\centering\includegraphics[width=\gsize]{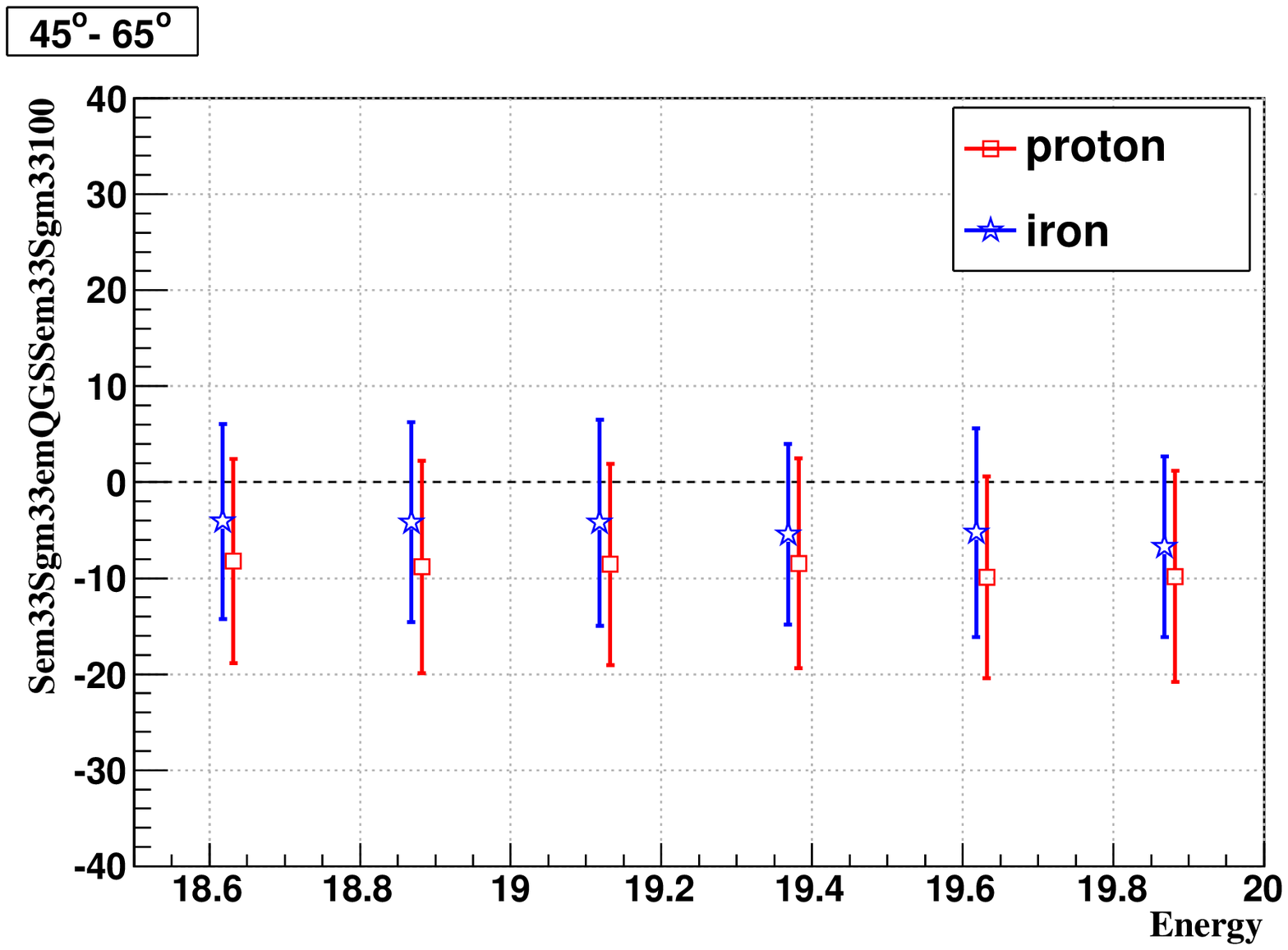}
\centering\includegraphics[width=\gsize]{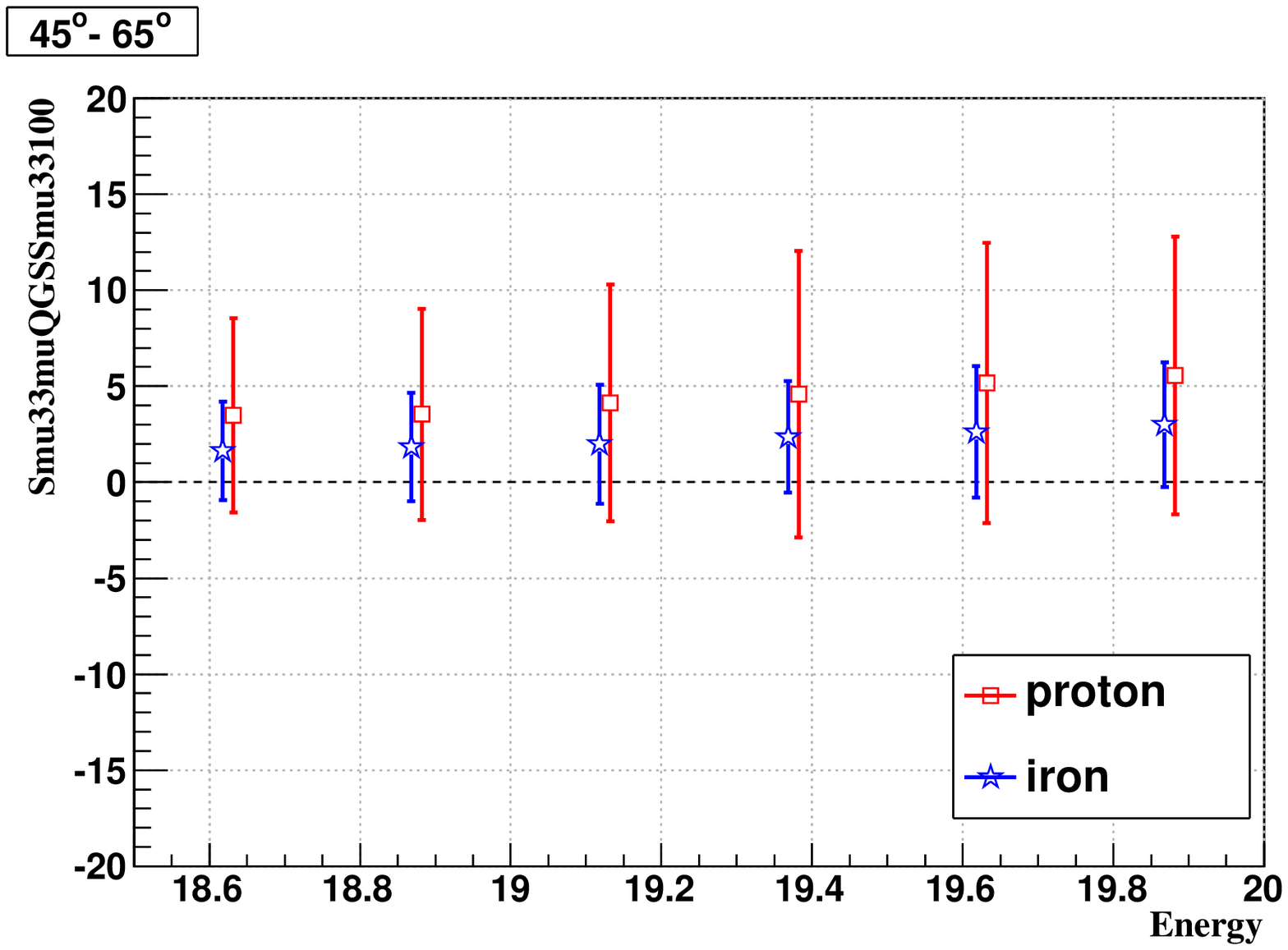}
\caption{Means and RMS of distributions of relative difference between
  MC simulated EM (top) and muon (bottom) EPOS~1.99 signals and
  signals obtained from the fit~(\ref{eq:smucos}) with parameters for
  QGSJET~II.}
\label{fig:diffcosQE}
\end{figure}

In Fig.~\ref{fig:diffcosQE} we present the result of application of
formula~(\ref{eq:smucos}) with the fit parameters for QGSJET~II to the
dataset simulated with EPOS~1.99. The increasing role of the EM halo,
that brings to almost the same scaling of total and muon
signals~\cite{ya_icrc2011}, and the similarity of \sigrat\ behavior on
\xmaxv\ allow to derive muon signal from EPOS~1.99 simulations with
errors below 6\% for protons and 3\% for iron showers applying
formula for QGSJET~II (and vice versa).  It is seen that with the
increase of the energy interaction model invariance becomes more
violated due to the increasing fraction of $\pi^0$ EM component
arriving at the observation level.

\section{Determination of the absolute signal scaling factor}

\begin{figure}[thb]
\centering\includegraphics[width=\gsize]{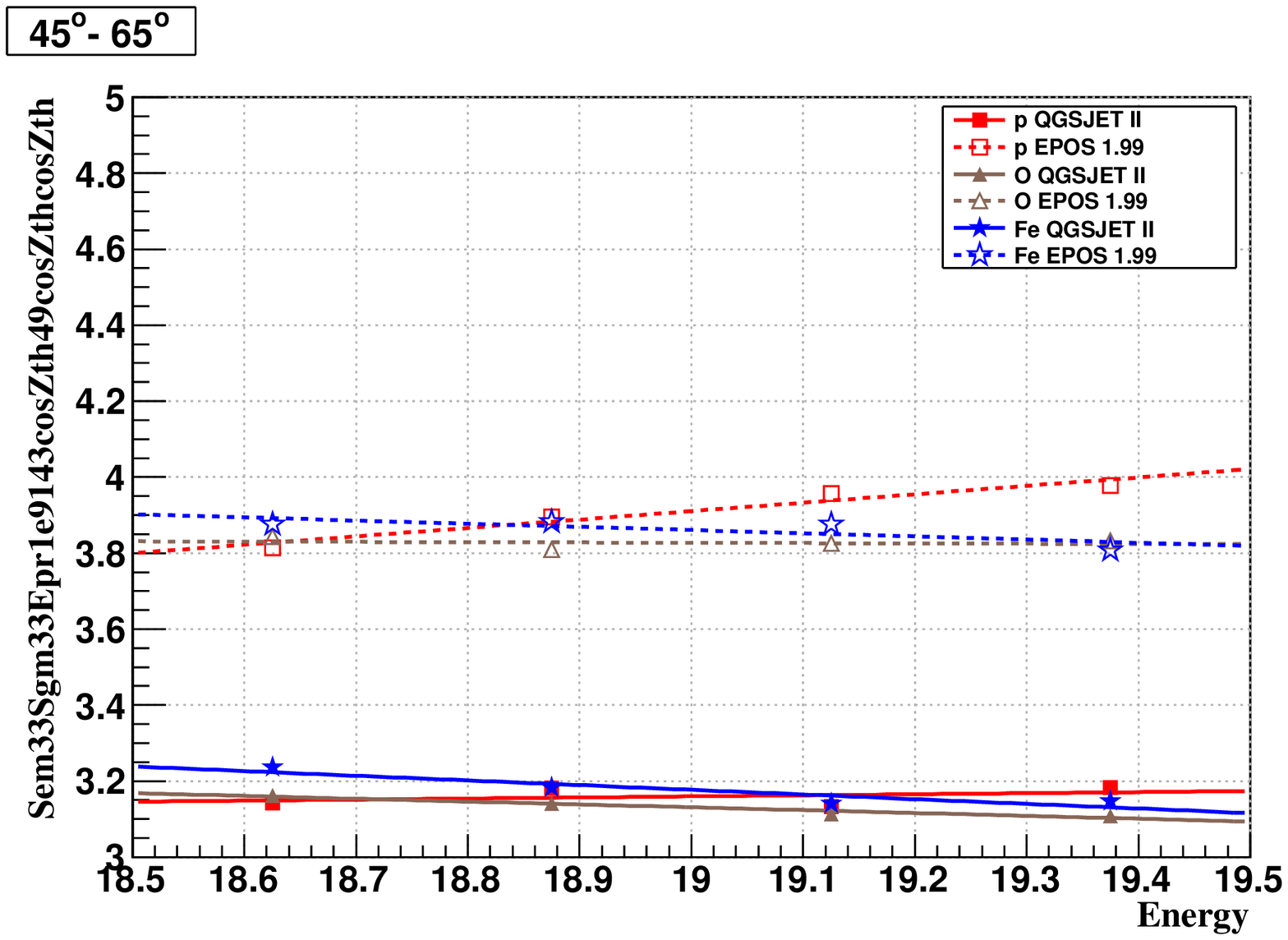}
\centering\includegraphics[width=\gsize]{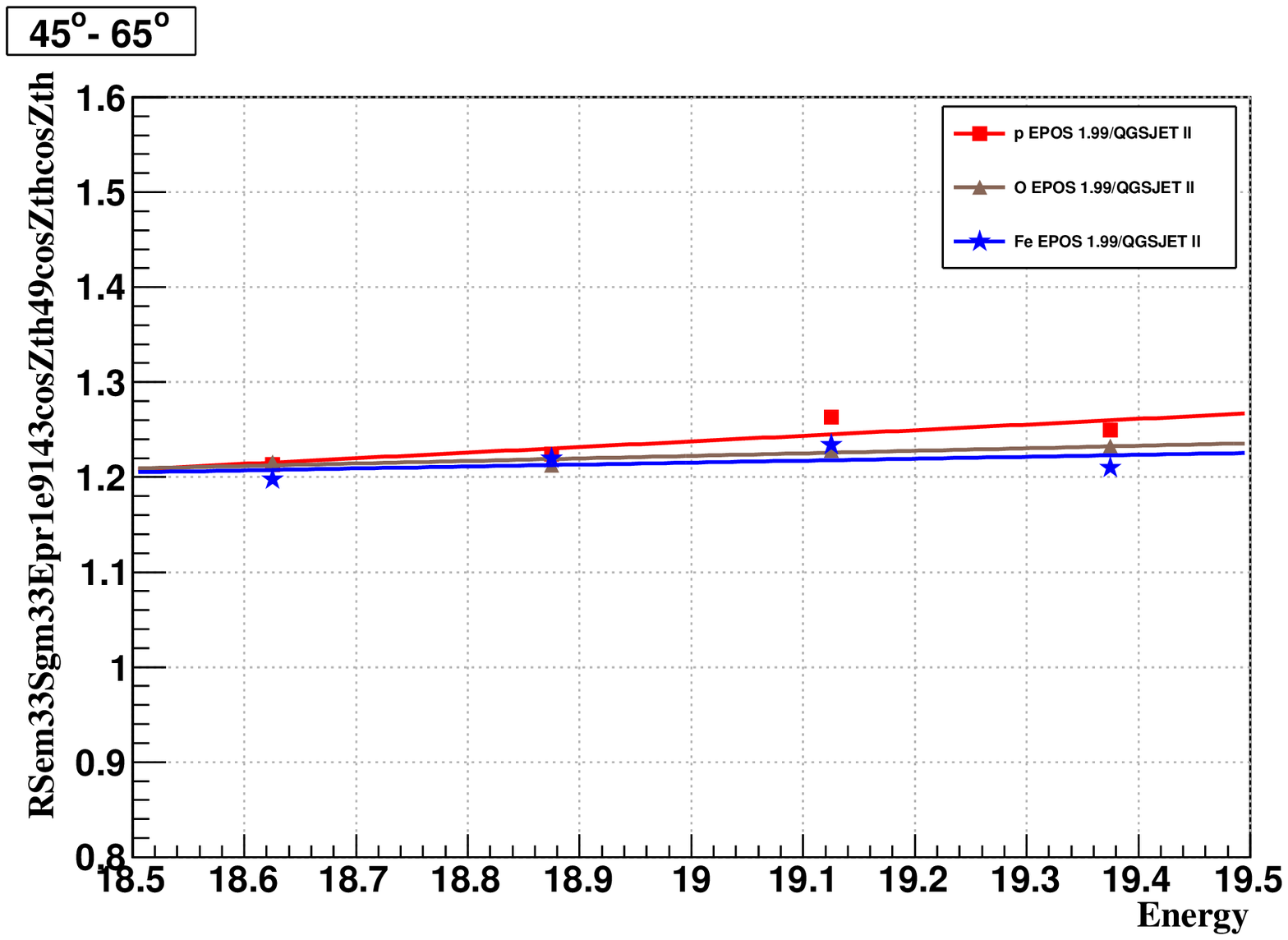}
\caption{Top: EM signals for QGSJET~II (closed symbols) and EPOS~1.99
  (open symbols) normalized by primary energy and
  $f(\theta)=1-4.3\cos(\theta)+4.9\cos^2(\theta)$. Bottom: ratio of the
  EPOS~1.99 to QGSJET~II EM signals, mean value is
  1.23.}
\label{fig:semnorm}
\end{figure}

\begin{figure}[thb]
\centering\includegraphics[width=\gsize]{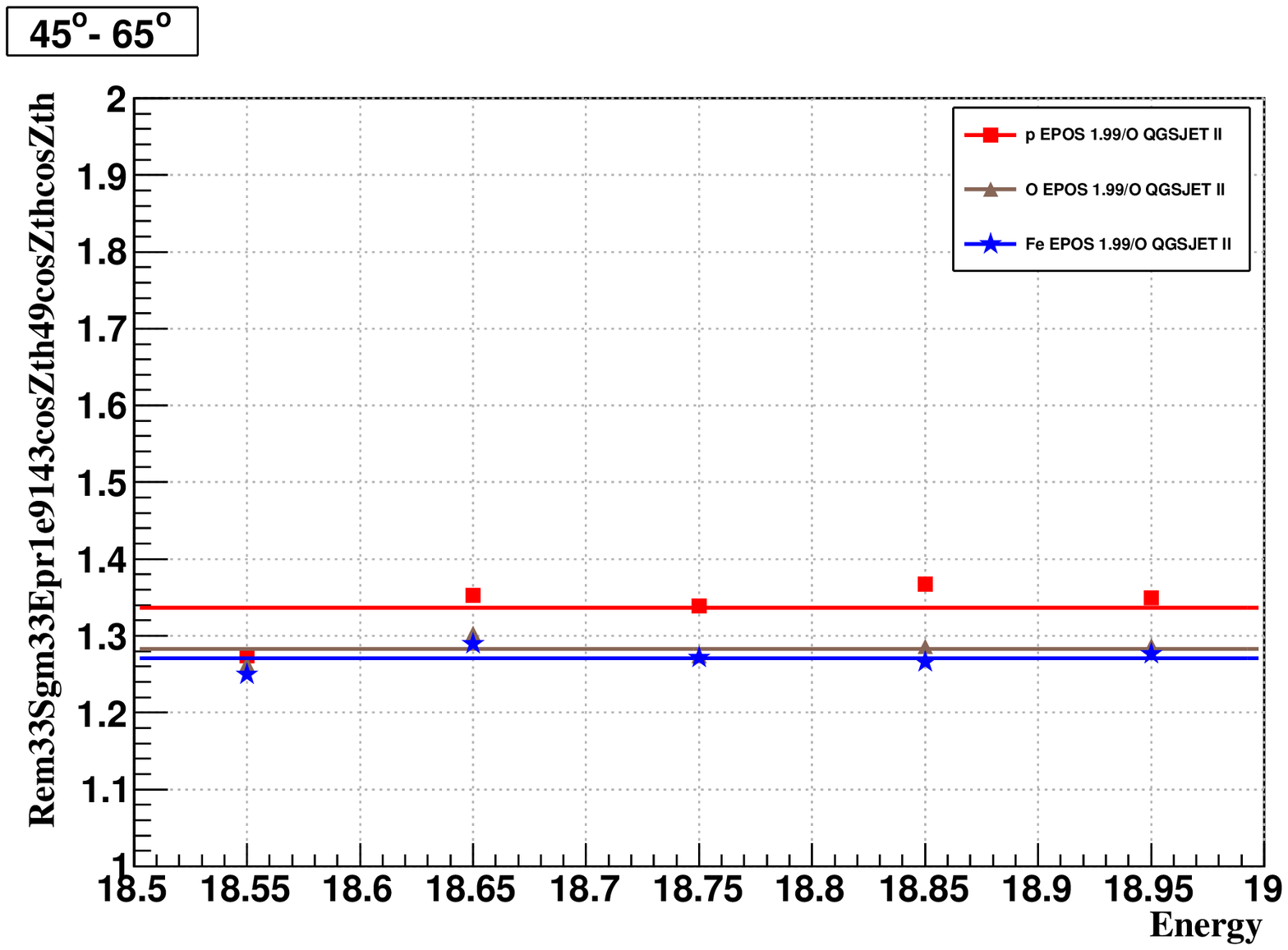}
\centering\includegraphics[width=\gsize]{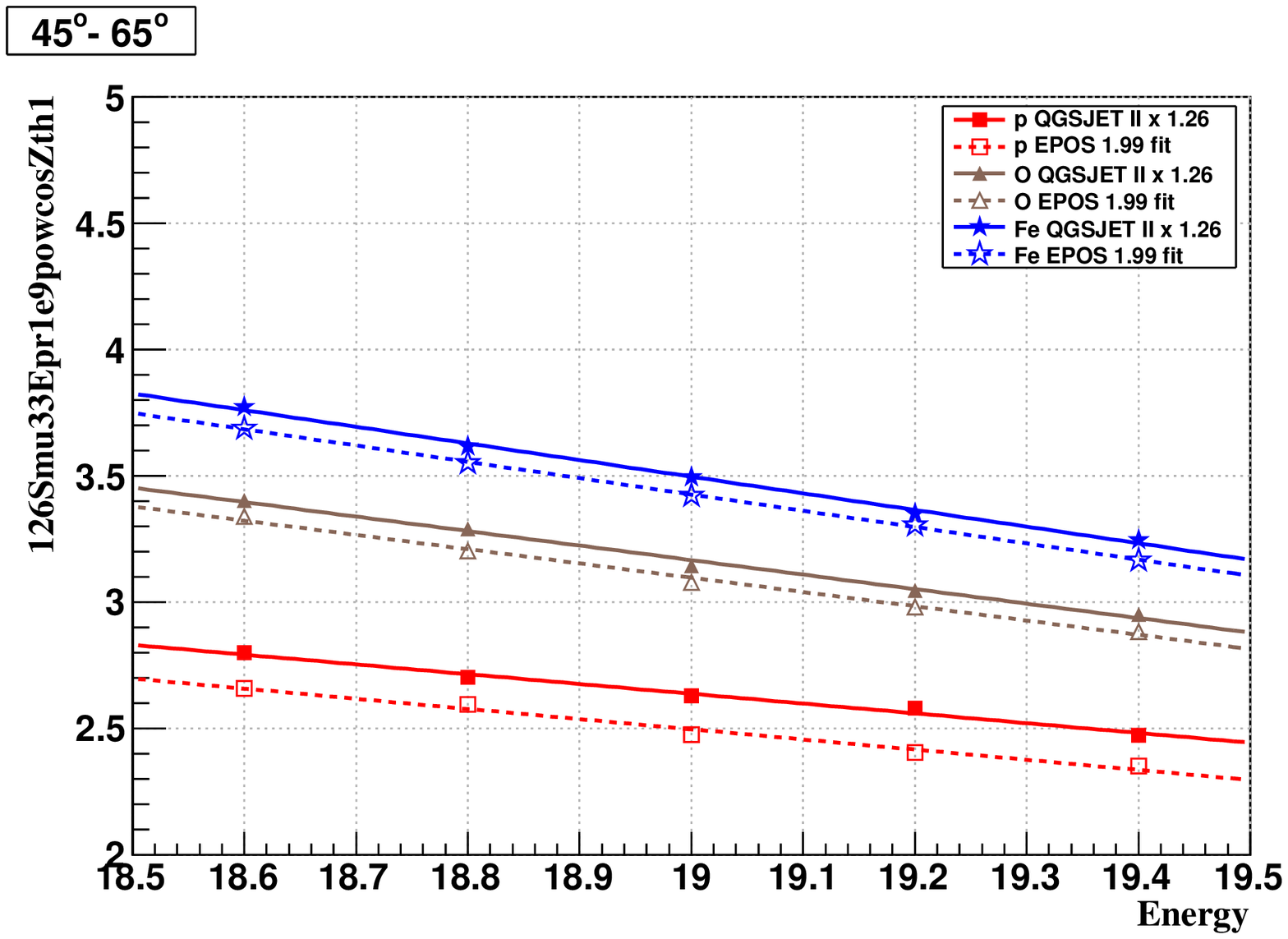}
\caption{Top: ratio of the parametrized EPOS~1.99 EM signals to the
  EM signal for QGSJET~II oxygen for energies below $10^{19}$
  eV fitted with flat line. Mean values are 1.34, 1.28 and 1.27 for
  $p$, O and Fe correspondingly. Bottom: scaled by 1.26 QGSJET~II muon
  signals and muon signals obtained from EPOS~1.99 simulated dataset
  with~(\ref{eq:smucos}) and coefficients for
  QGSJET~II.}
\label{fig:smuscaledfitted}
\end{figure}

\begin{figure}[htb]
\centering\includegraphics[width=\gsize]{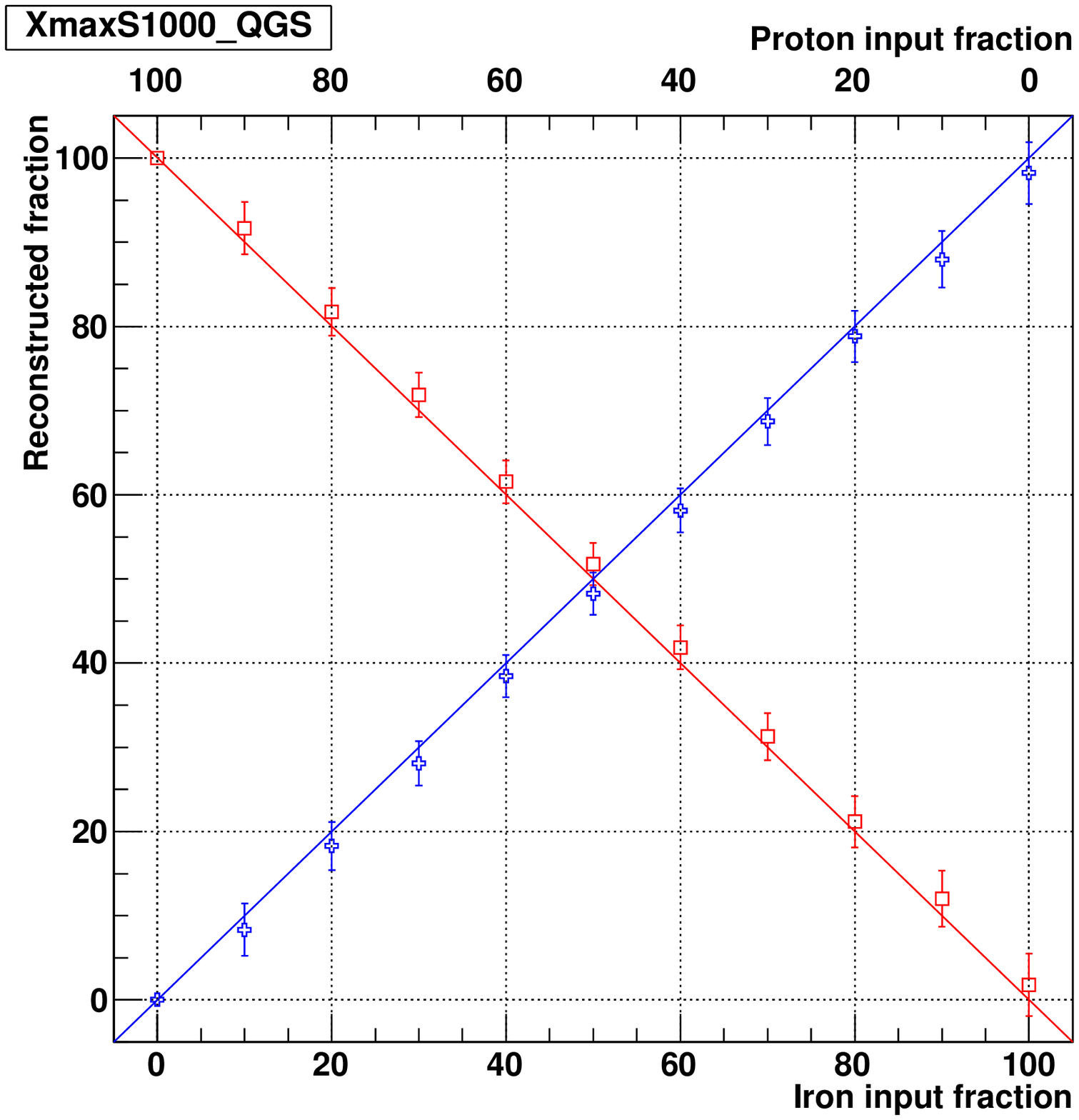}
\centering\includegraphics[width=\gsize]{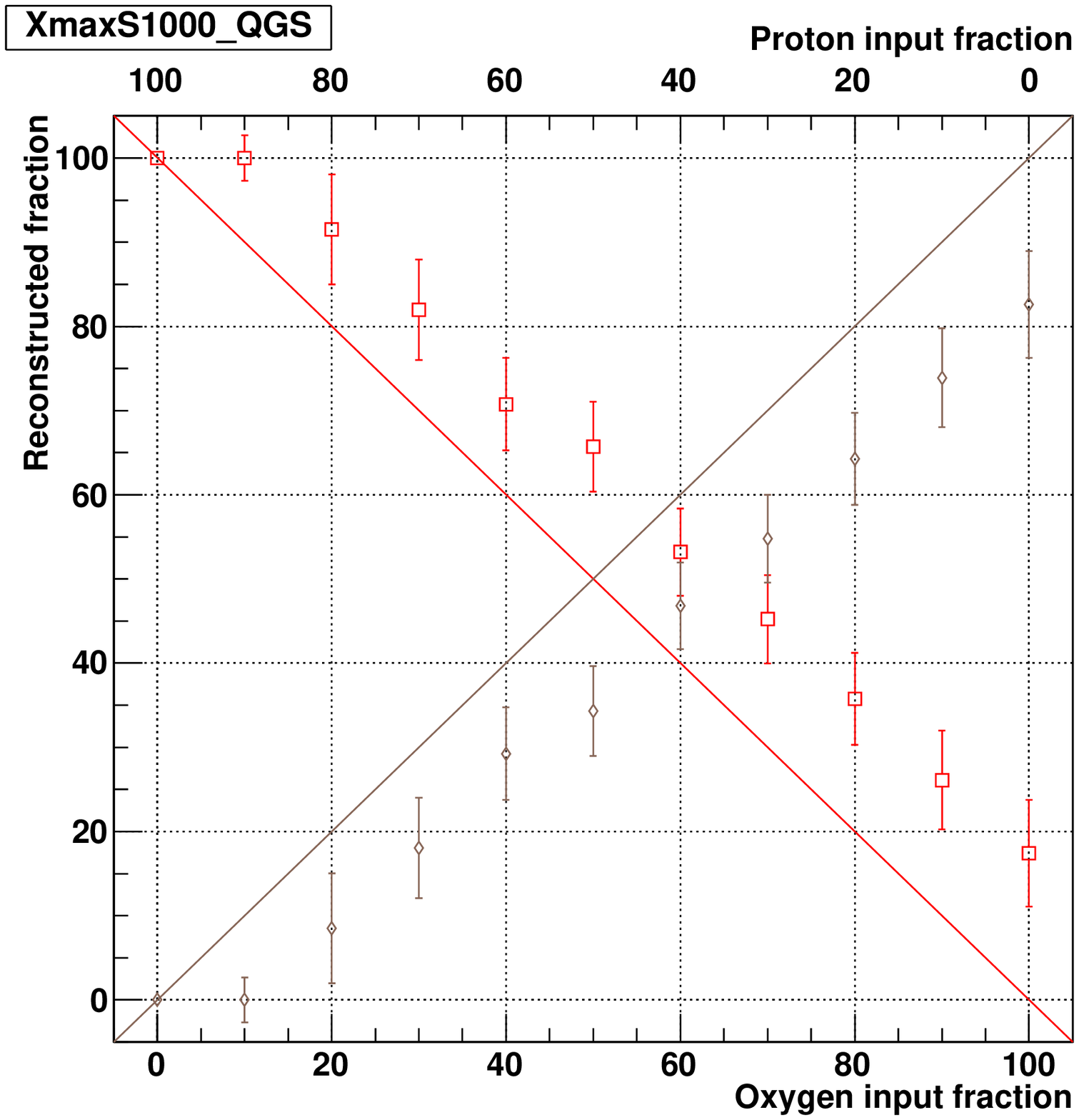}
\caption{Top: reconstructed with QGSJET~II and MTA on the basis of
  Fisher's variables distributions proton (red squares) and iron (blue
  crosses) abundances in the EPOS~1.99 samples with known primaries
  content. Bottom: same but for proton--oxygen (brown diamonds)
  mixture. Lines mark the true primary fractions. $\logen18.90-19.10$}
\label{fig:mass}
\end{figure}

The muon signal is one of the most powerful shower characteristics for
the mass composition analysis. The procedure of determination of the
muon signal proposed in the previous section provides the way to get
the muon signal from hybrid data in Auger-like experiments, but due to
well-known problem of muon deficit in Monte-Carlo (MC) shower
simulation
codes~\cite{abuzayyad_mass2000,engel_icrc2007,ave_munum_2007} the
retrieved signal can not be used in mass composition studies. The
latter could become possible only if one would be able to find
absolute signal scaling factor
$S(p,\,\mathrm{real\ data})/S(p,\,\mathrm{MC})$. We would like to
propose a possible way to solve this problem using as a fake data
EPOS~1.99 simulations and QGSJET~II as a test interaction model. Of
course, for our fake data produced with EPOS~1.99 we know precisely
the total, muon and EM signals and this will help us to estimate the
accuracy of the proposed procedure.

Since primary mass composition is unknown the only way to determine
the scaling factor is to use mass composition independent shower
characteristics and the only appropriate candidate to this role is the
EM signal. In Fig.~\ref{fig:semnorm} one can see its mass independence
for QGSJET~II and EPOS~1.99 (the signals are divided by
$f(\theta)=1-4.3\cos(\theta)+4.9\cos^2(\theta)$ to reduce spread
caused by the spread of zenith angles) in the energy range
\erange{18.5}{19.5}. At the bottom it is shown the ratio of the
EPOS~1.99 EM signals to QGSJET~II ones and the average of this ratio
is equal to 1.23. Due to increasing role of the EM halo in the
considered angular range \arange{45}{65} \sigrat\ becomes less
dependent on the hadronic interaction properties~\cite{ya_icrc2011}
providing the following approximate equality which holds true for any
primary nuclei ($p$, O, Fe etc.):
\begin{equation}
\label{eq:allscales}
\frac{\smillam{EPOS}}{\smillam{QGS}}\approx\frac{\smum{EPOS}}{\smum{QGS}}\approx\frac{\semm{EPOS}}{\semm{QGS}}.
\end{equation}
The ratios of the total and muon signals confirm these relation with
quite good accuracy: their average values for all primaries (1.25 and
1.26) are quite the same as the value for the EM signals. The ratios
of the total and muon signals remain constant across the entire
considered energy range.

Now it remains only to try to scale the QGSJET~II model predictions
using EPOS~1.99 set as a fake data. To find EM and muon signals in the
`data' one should simply apply the parametrization~(\ref{eq:smucos})
with coefficient for QGSJET~II to EPOS~1.99 simulations and to find
the ratio of the EM signal from the `data' to the QGSJET~II one. From
Fig.~\ref{fig:diffcosQE} one can see that the error on the EM signal
extracted from EPOS~1.99 can reach 10\% for proton primaries and 6\%
for iron. One can see also that with the increase of the energy the
error in determination of the scaling factor should increase due to
increasing fraction of the $\pi^0$ EM component arriving at ground
level. Hence, it is reasonable to limit the scaling by the energy
range \erange{18.5}{19.0} and in this case one gets the scaling
factors equal to 1.34, 1.28 and 1.27 for $p$, O and Fe correspondingly
(Fig.~\ref{fig:smuscaledfitted}, normalization is done in respect to
QGSJET~II oxygen). The true ratio of the muon signals is 1.26 and in
case of pure iron or mixed primary composition after scaling both
models should give very close predictions of the muon signals, for
pure primary proton flux the discrepancy between the true EPOS~1.99
signals and scaled QGSJET~II signals can exceed 5\%. Evidently, this
error is due to deep proton showers and if one applies
$\xmaxv<500$~\gsm\ cut only during scaling procedure this will affect
almost exclusively proton showers and will change the scaling factor
for them to 1.26. Hence, using the cut $\xmaxv<500$~\gsm\ the scaling
factor that one gets will be within $1.26-1.28$ range independently on
the mass composition, in very good agreement with the true muon
signals ratio. In Fig.~\ref{fig:smuscaledfitted} the muon signal
obtained from EPOS~1.99 `data' with~(\ref{eq:smucos}) and parameters
for QGSJET~II is compared with the scaled by 1.26 QGSJET~II muon
signal. One can see that after the entire procedure one gets quite
consistent picture with the tendency to underestimation of the `true'
primary mass, since e.g. the `true' proton signals retrieved with the
fit is lower than scaled QGSJET~II signals to which the comparison
should be done.

We have reconstructed a mass composition of proton-oxygen and
proton-iron mixtures prepared with EPOS~1.99 using scaled by 1.26
QGSJET~II model signals. The muon signal from EPOS~1.99 samples was
retrieved with the use of the fit~(\ref{eq:smucos}) with parameters
for QGSJET~II. To discriminate primaries $(\xmax,\,
\smu^\mathrm{fit}/(E\xmaxv))$ variables have been used and the
approach has been the same as in~\cite{ya_Lodz}, i.e. with consequent
application of the Fisher's discriminant analysis and Multiparametric
Topological Analysis (MTA)~\cite{durso_mta_2008}. From
Fig.~\ref{fig:mass} one can see that for proton-iron mixtures the
method gives excellent results, while for proton-oxygen mixture the
reconstructed composition is lighter than the original one and errors
grow with the increase of the oxygen fraction from 10 to 17\%. Let us
note that these errors are almost completely due to errors in muon
signal and scaling factor determination (see
Fig.~\ref{fig:smuscaledfitted}), while accuracy provided by MTA itself
is better than 2\%. More precise results for $p$\,--\,Fe samples are
explained by very good separation of the Fisher's variable
distributions for these primaries~\cite{ya_Lodz} and small errors on
the scaling factor do not influence significantly the events
misclassification rate. For the scaling factor of 1.28 the accuracy of
reconstruction is $2-4$\% and $15-20$\% for $p$\,--\,Fe and $p$\,--\,O
mixtures correspondingly.

\section*{Conclusions}
In this paper we have demonstrated that the universality of
\sigrat\ ratio in \arange{45}{65} angular range in respect to the
interaction model properties allows to get the muon signal from hybrid
data with accuracy of $3-5$\%. The application of this approach to the
data of the Pierre Auger Observatory can be found
elsewhere~\cite{auger_nmu_icrc2011}. Further, using the independence
of the EM signal on the primary mass we have proposed a procedure
giving a possibility to find absolute scaling factor between real data
and MC simulated signals. Using EPOS~1.99 simulations as a fake data
we have found a scaling factor for QGSJET~II signals with an accuracy
of few percents. Application of the scaled QGSJET~II muon signals
allowed to reconstruct mass composition of the samples prepared with
EPOS~1.99 with errors below 4\% for proton-iron mixtures, while for
proton-oxygen ones the accuracy is around $10-20$\%. Hence, the use of
the both models in the proposed way for reconstruction of the real
primary mass composition will give closely agreeing results. The
preference to results obtained with one of the models can be given on
the basis of the comparison with measurements of \xmax\ and
\smu\ distributions.

\subsubsection*{Acknowledgments}
We are very grateful to Maximo Ave and Fabian Schmidt for kind
permission to use their GEANT~4 lookup tables in our calculations of
signal from different particles in Auger water Cherenkov detectors.

%\bibliographystyle{/home/yushkov/bin/bibtex/ICRC2005} 
%\bibliography{/home/yushkov/bin/bibtex/my_rus}

\end{document}